\documentclass[a4paper,aps,11pt,preprint]{revtex4}
\usepackage{graphicx}
\usepackage{hyperref,amsmath,amssymb,textcomp}
\begin{document}

\title{Universal main magnetic focus ion source: A new tool for laboratory research of astrophysics and Tokamak microplasma}

\author{V.P. Ovsyannikov}\thanks{
URL: \url{http://mamfis.net/ovsyannikov.html} }
\affiliation{ MaMFIS Group, Hochschulstr. 13, D-01069  Dresden, Germany}

\author{A.V. Nefiodov}\thanks{E-mail: anef@thd.pnpi.spb.ru}
\affiliation{Petersburg Nuclear Physics Institute, 188300 Gatchina, St.~Petersburg, Russia}

\author{A.A.  Levin}\thanks{E-mail: aalevin63@gmail.com}
\affiliation{MaMFIS Laboratory, Shatelen Street 26A, 194021 St.~Petersburg, Russia}

\widetext

\begin{abstract}
A novel room-temperature ion source for the production of  atomic ions in electron beam within wide ranges of electron energy  and current density is developed. The device can operate both as conventional Electron Beam Ion Source/Trap (EBIS/T) and novel Main Magnetic Focus Ion Source.  The ion source is suitable for generation of  the low-, medium- and high-density microplasma in steady state, which can be employed for investigation of a wide range of physical problems in ordinary university laboratory, in particular, for  microplasma simulations relevant to astrophysics and ITER reactor. For the electron beam characterized by the incident energy $E_e = 10$~keV, the current density $j_e \sim 20$~kA/cm$^2$ and the number density $n_e \sim 2 \times 10^{13}$~cm$^{-3}$ were achieved experimentally. For $E_e \sim 60$~keV, the value of electron number density $n_e \sim 10^{14}$~cm$^{-3}$ is feasible.  The efficiency of  the novel ion source for laboratory astrophysics significantly exceeds that of  other existing warm  and superconducting EBITs.  A problem of  the K-shell electron ionization  of  heavy and superheavy elements  is also discussed.
\end{abstract}
\maketitle

\section{Electron beam ion sources and traps}

Nowadays, highly charged ions are not only the object of  scientific investigation in atomic and plasma physics, laboratory astrophysics  etc., but also the technological tool in the fields of accelerator techniques, ion microscopy, surface machining  on the nanoscale, ion therapy and elsewhere. The specific charge state of ions is crucial in a number of applications. In the case of  the successive multiple ionization of  ions by electron impact,  the charge states $+q$ are determined by the ionization factor $j_e \tau$, where $j_e$ is the electron beam current density and $\tau$ is the duration of  ion bombardment by incident electrons (ionization time).   The latter is bounded above by the time of  transition of  plasma into a steady state.  It is easy to estimate that the production of ions in high-charge states necessitates the dense electron  beam and significant ionization time.   Accordingly, there is inevitable necessity to employ a trap for confinement of ions.

\begin{figure}[b]
\begin{minipage}{18pc}
\includegraphics[width=18pc]{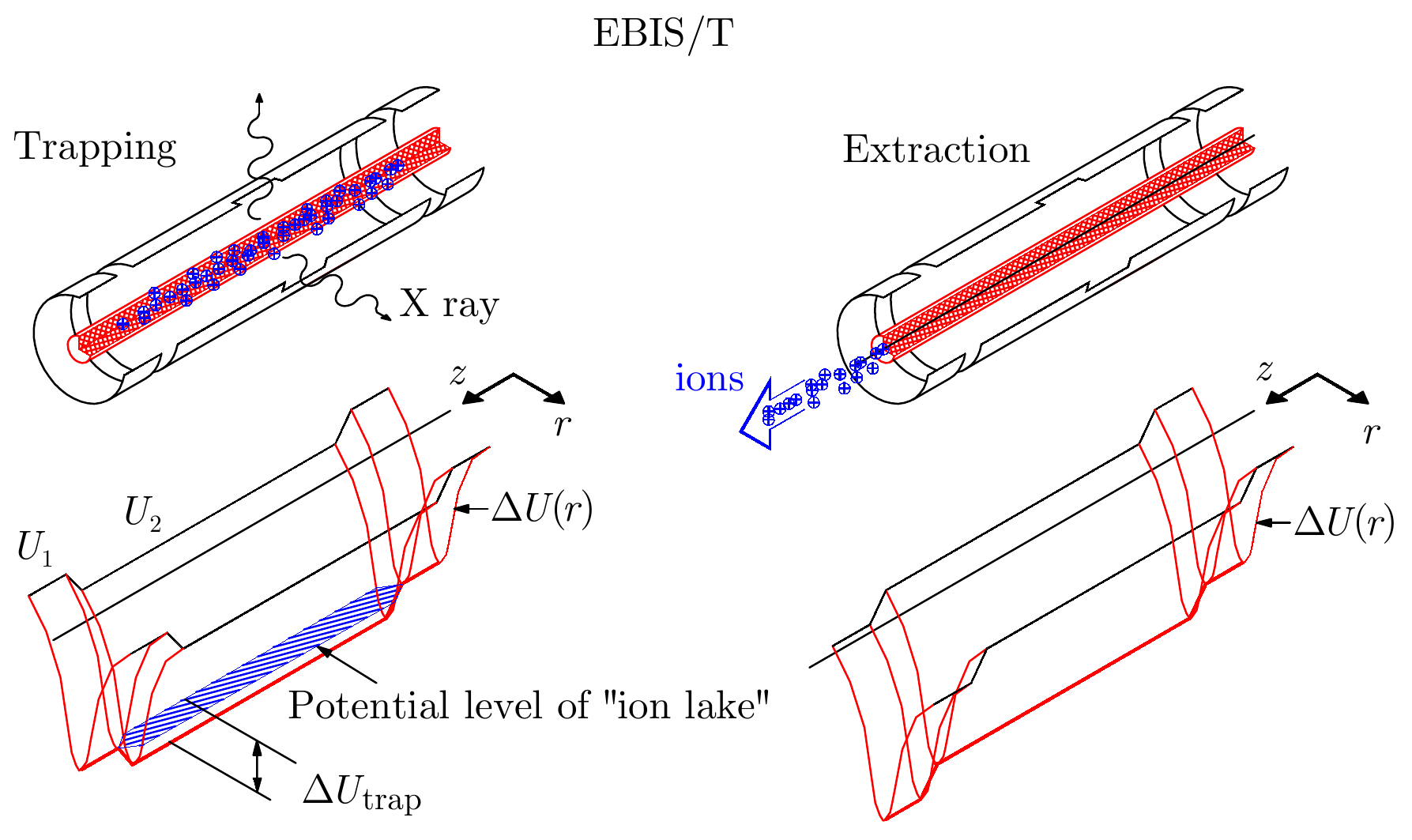}
\caption{\label{fig1} Principal scheme of the  EBIS/T. The $z$ axis is  directed  along the electron beam, $U_1$ and  $U_2$  are potentials of  the adjacent sections and $\Delta U(r)$ is the radial sag of potential in space of  the drift tube. }
\end{minipage}\hspace{2pc}%
\begin{minipage}{18pc}
\includegraphics[width=18pc]{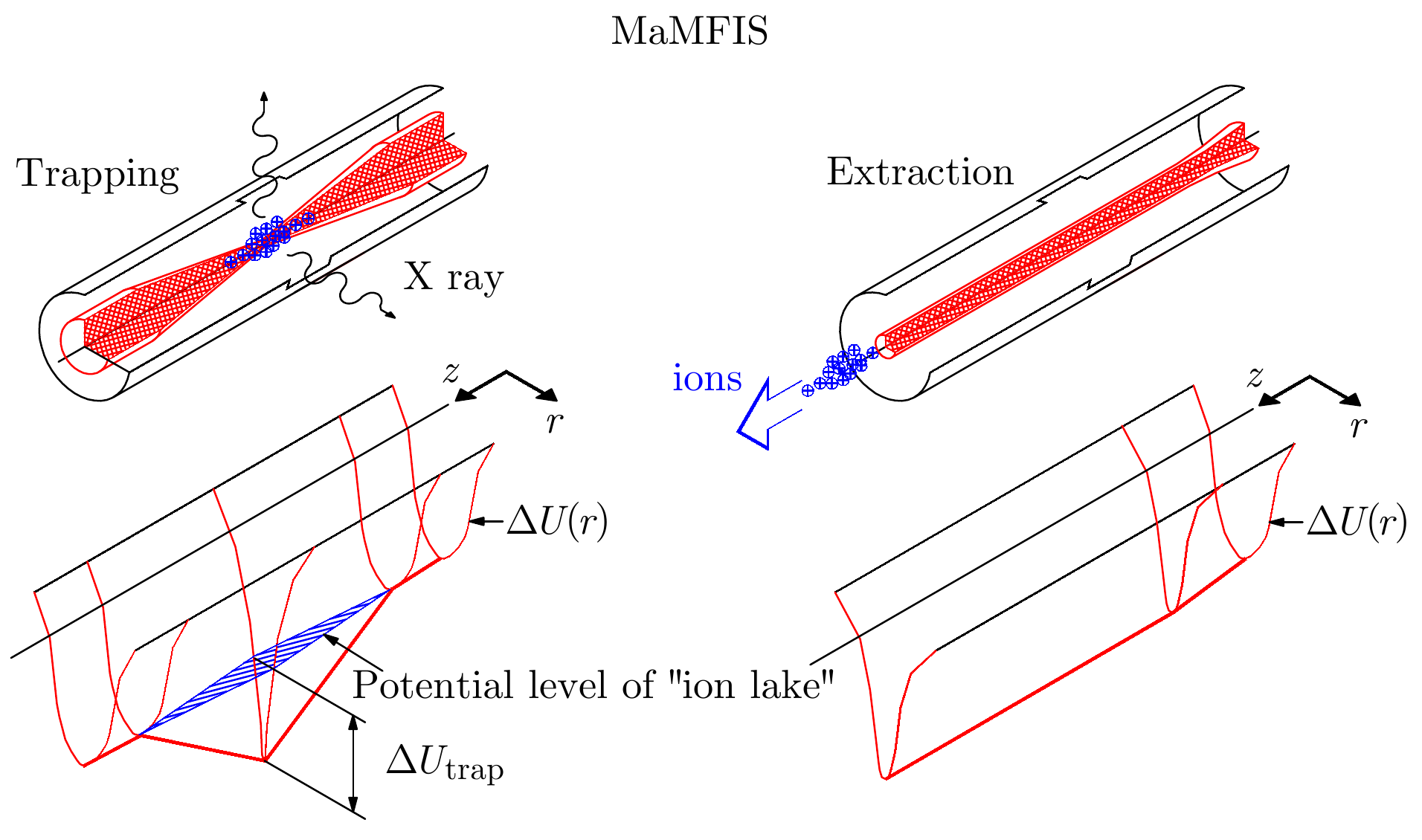}
\caption{\label{fig2} Principal  scheme of  the MaMFIS. The notations are similar to those in Fig.~\ref{fig1}, while 
potential ion trap is formed by local structure of the electron beam itself and  described by Eq.~\eqref{eq1}.}
\vspace{1.1pc}%
\end{minipage} 
\end{figure}

A trap formed by the axially symmetric smooth electron beam, which propagates  through the cylindrical drift tube consisting of a few (at least three) sections with positive potentials applied to the edge sections, was suggested by E.D.~Donets in 1967  \cite{1}. The method was realized in a device, which was  named the Electron Beam Ion Source (EBIS).  The ions are confined  by volume charge of the electron beam and by external potentials  at both  ends of the drift tube. The depth of  ion trap is given by difference of  the maximum and minimum potentials: $\Delta U_\mathrm{trap}= U_1 - U_2$  (see  Fig.~\ref{fig1}). The axially symmetric magnetic field  is also applied to compress the electron beam and hinder the radial escape of ions from the potential well. 
 
The original ion sources  were developed for injection complexes of accelerators \cite{2}. The devices were intended to provide the high-intensity ion beams.  This required creation of the electron-optical systems with electron beam of  about 1~m in  length and characterized by current of  a few Ampers \cite{3}.  For the ion sources employed in real  accelerators,  the electron current density falls in the range of 100--500~A/cm$^2$  \cite{4}.  Such values of $j_e$  are not sufficient for the production of  highly charged ions of  heavy elements of the periodic table.  In 1986, the Electron Beam Ion Trap (EBIT) was constructed at the Lawrence Livermore National Laboratory  (LLNL)  for the purposes of X-ray spectroscopy  \cite{5}. Physically, the EBIT is similar to the EBIS operating in the trapping regime.  However, the distinguishing feature of such device is the small length of ion trap of about 2 cm, what allowed one to increase significantly the current density $j_e$ up to 5~kA/cm$^2$ and to ionize heavy elements up to uranium \cite{6}. The classical Livermore EBIT with the electron beam energy $E_e$ of  up to 30 keV  is now widely replicated around the globe. The devices similar to the high-energy  SuperEBIT  aiming at the production of  highly charged  heavy ions  were also constructed  in Germany, Japan and China \cite{7,8,9}.  All the devices mentioned above employ the cryogenic engineering and superconducting magnetic focusing systems. These   apparatuses remain to be very complicated and expensive to use  and, accordingly, are  available to large laboratories and national reseach centers only.

In 1999, a warm EBIT with electron beam focused by the radially magnetized permanent magnets was successfully tested  in Dresden. A whole family of ion sources and traps named  the  Dresden EBIS/Ts  was developed on the basis of this technology together with  standard methods of obtaining the ultra-high vacuum \cite{10}.  At present, the devices operate at different universities and research centers.  The important feature of  the Dresden EBIS/Ts is  a room working temperature, which significantly decreases the exploitation costs. However, although the electron beam energy achieved the values of about 10~keV, the current density $j_e$ of  these devices does not exceed  0.3~kA/cm$^2$.

\begin{figure}[b]
\begin{center}
\begin{minipage}{16.55pc}
\includegraphics[width=16.55pc,angle=0,keepaspectratio,clip]{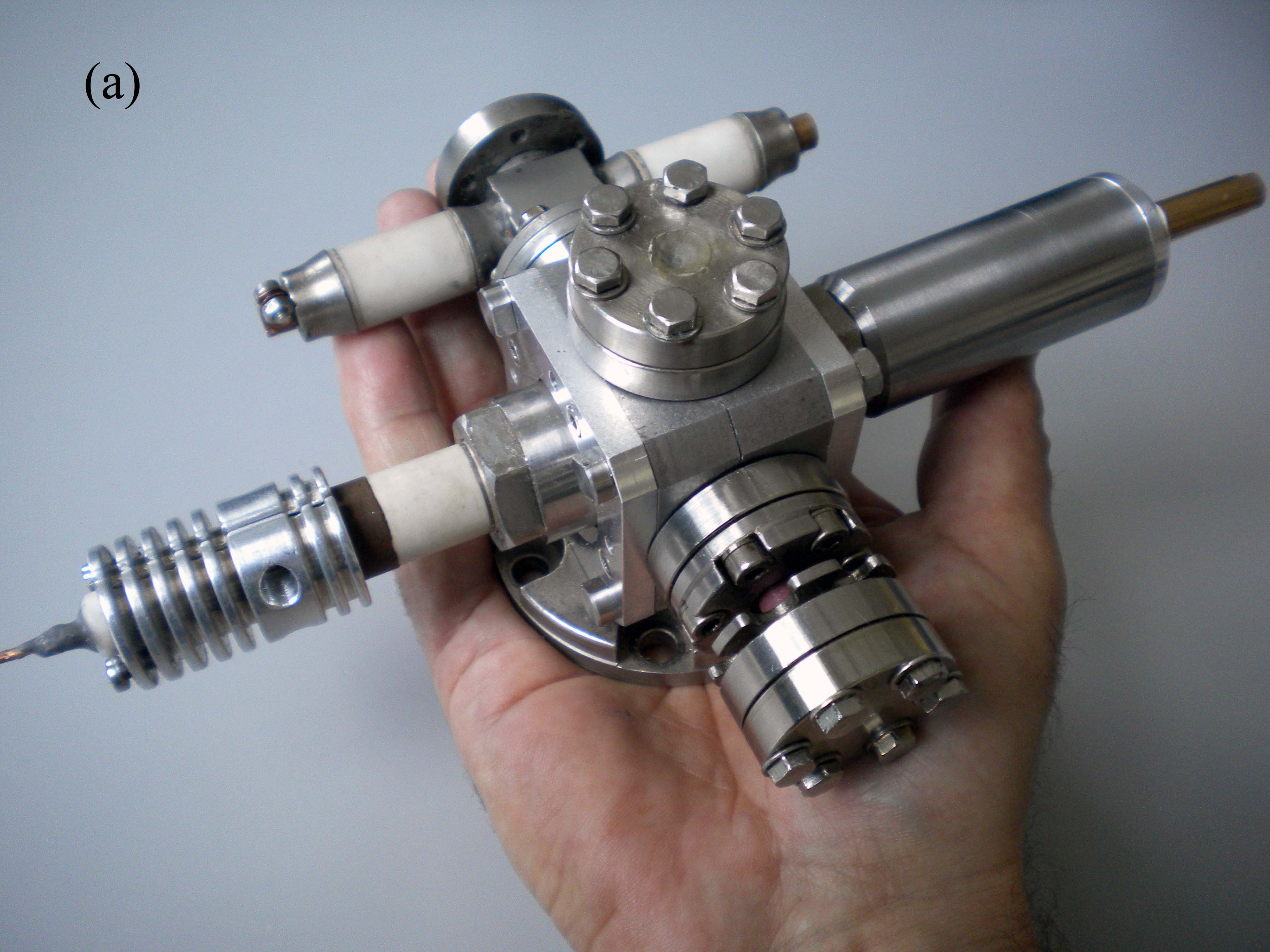}
\end{minipage}\hspace{0.7pc}%
\begin{minipage}{9.8pc}
\includegraphics[width=9.8pc,angle=0,keepaspectratio,clip]{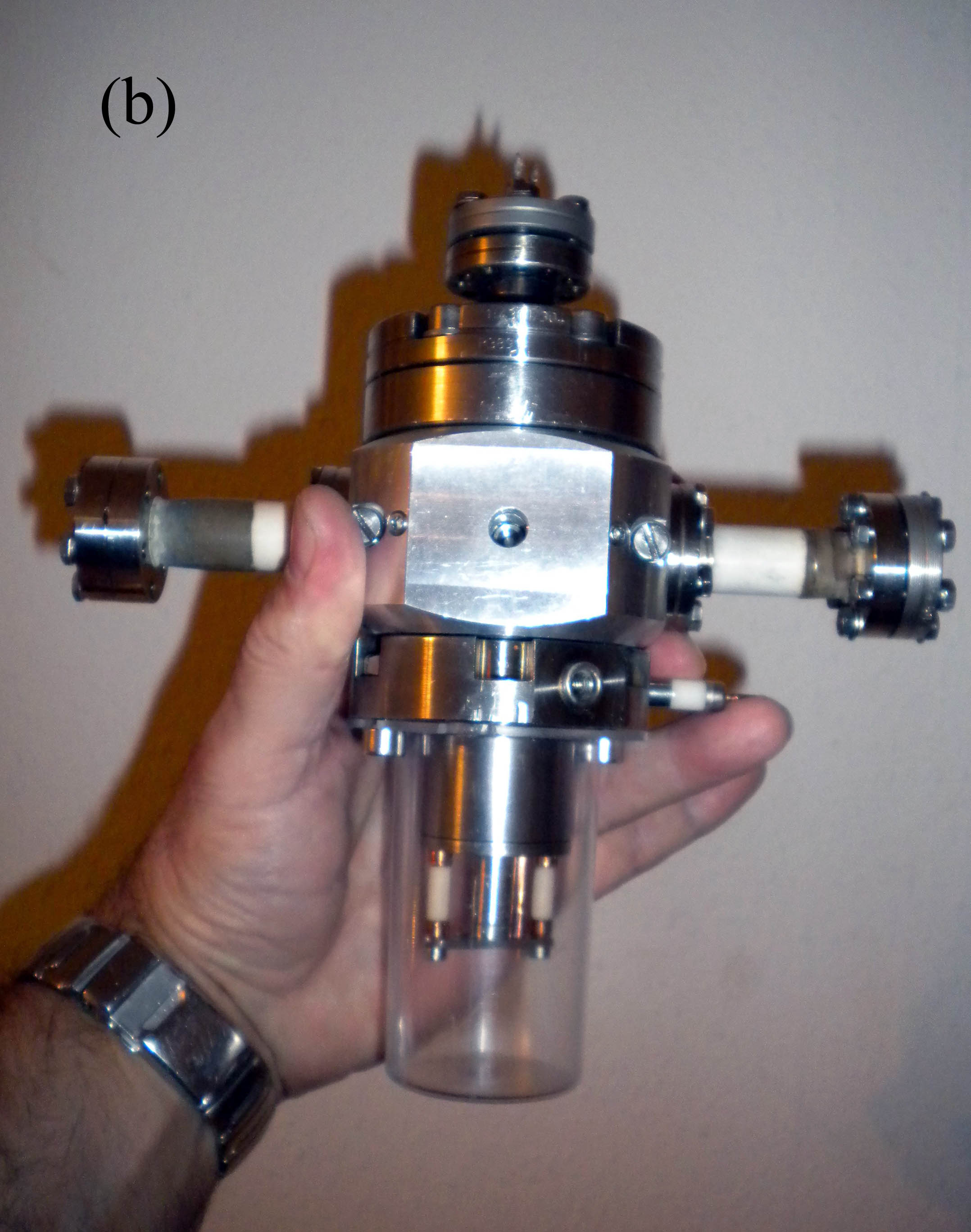}
\end{minipage} \hspace{0.4pc}%
\begin{minipage}{9.2pc}
\includegraphics[width=9.2pc,angle=0,keepaspectratio,clip]{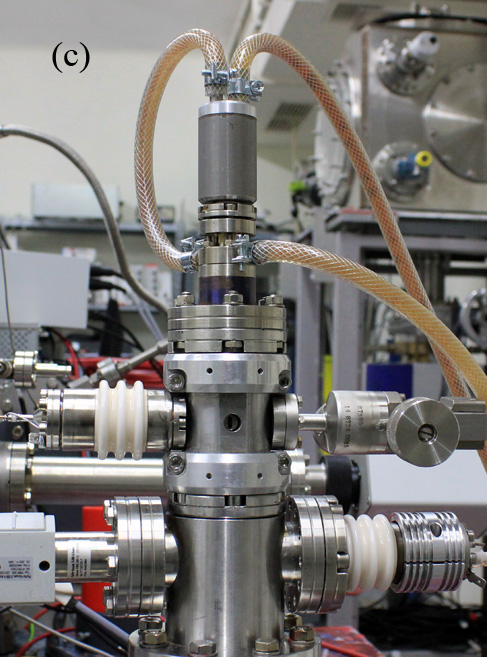}
\end{minipage} 
\caption{\label{fig3} (a)~MaMFIS-4, (b)~MaMFIS-10 and (c)~MaMFIS-30.}
\end{center}
\end{figure}

\section{Main magnetic focus ion source}

The electron current density is the key for the production of highly charged ions. In  EBIS/Ts, the magnitude of $j_e$ for given values of the focusing magnetic field $B_c$ at cathode and the energy  of  electron beam is limited by  requirements of  the Brillouin focusing. In reality, the true density is even lower because of  the influence of  thermal velocities  \cite{11}.  A substantial increase in current density can be achieved in the local ion  trap formed by a rippled electron beam in a cylindrical  drift tube \cite{12,13,14}.  In this case, the electron beam is focused  by a thick magnetic lens in a sharp crossover, where $j_e$ can reach values of  up to 100 kA/cm$^2$. This principle is realized in the Main Magnetic Focus Ion Source (MaMFIS)  (see  Fig.~\ref{fig2}). The device operates at  room temperature due to the use of permanent magnets and standard vacuum techniques.

The depth of  local ion trap  is estimated  as
\begin{equation}\label{eq1}
\Delta U_\mathrm{trap}=  \frac{PU}{2 \pi \varepsilon_0 \sqrt{2 \eta}}  \ln \frac{r_\mathrm{max}}{r_\mathrm{min}}  .
\end{equation}
Here $r_\mathrm{max}$ and  $r_\mathrm{min}$ are the maximum and minimum values of radius of the electron beam, $\varepsilon_0$ is the permittivity of free space, $\eta = e/m$ is the electron charge-to-mass ratio, $U$ is the accelerating voltage, $P= I_{e}/U^{3/2}$ is the perveance of the electron beam and $I_{e}$ is the electron current. The  extraction of ions  from the source is realized by changing the shape of  electron beam. The rippled electron beam is transformed into a smooth flow with constant radius, when changing potential of the focusing electrode.  Although the ion trap  is just about 1 mm in  length, it is compensated by extremely high values of $j_e$ and small sizes  of the device.

\begin{figure}[t]
\begin{minipage}{18pc}
\includegraphics[width=18pc]{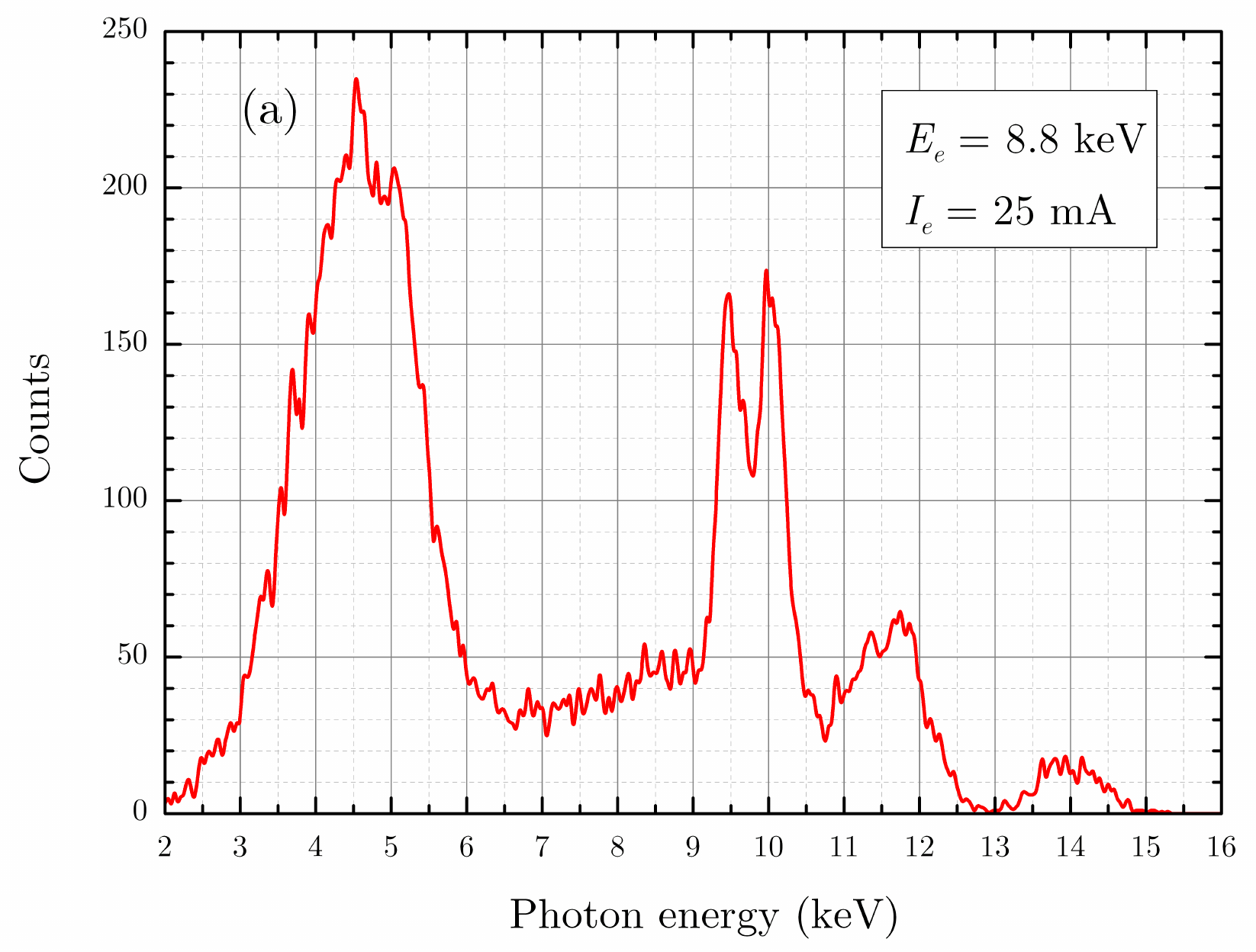}
\end{minipage}\hspace{2pc}%
\begin{minipage}{18pc}
\includegraphics[width=18pc]{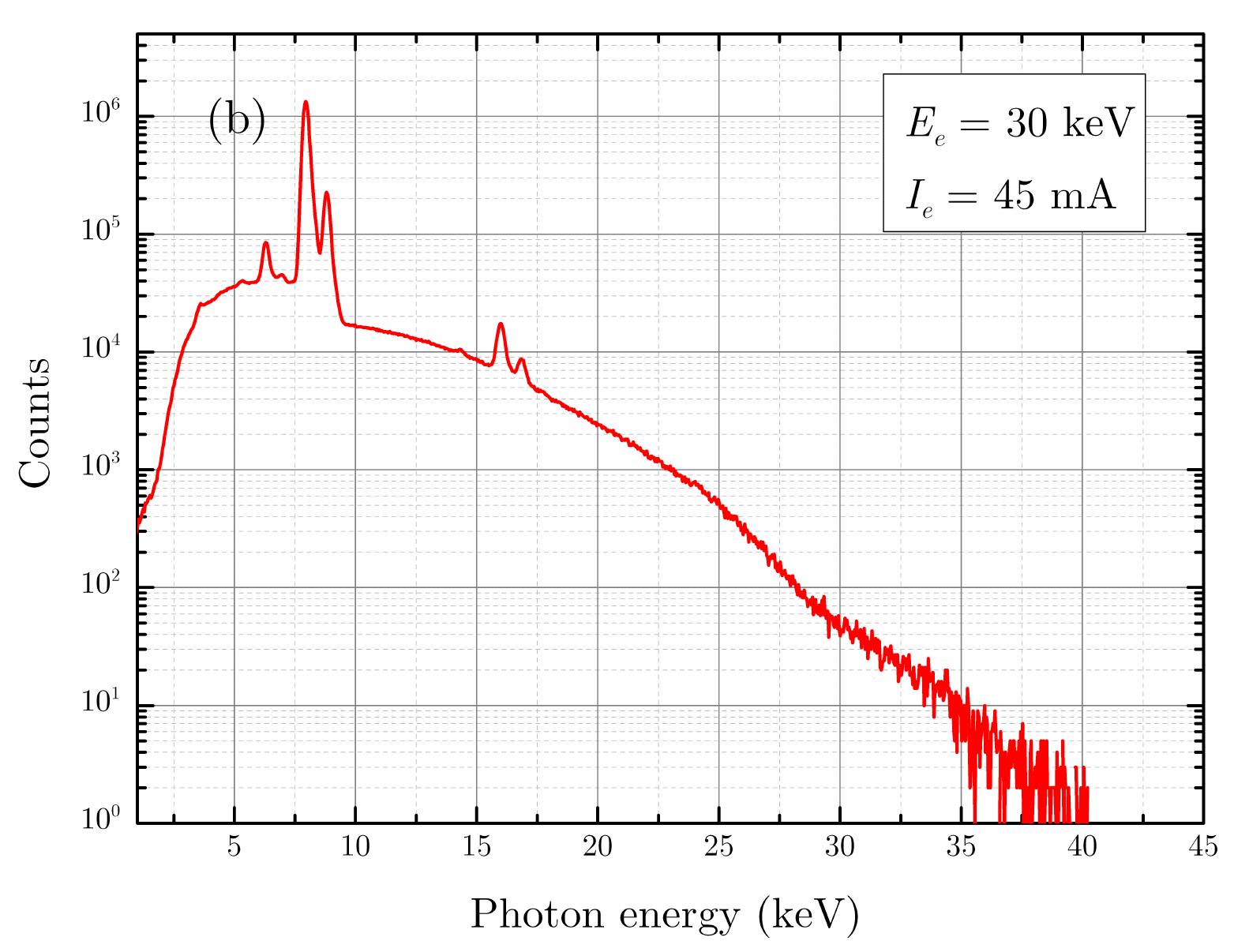}
\end{minipage} 
\caption{\label{fig4}  X-ray radiation of  Ir and  Ce ions from the local ion trap of  MaMFIS.}
\end{figure}

Presently, prototypes of  the MaMFIS with $E_e$ of  up to 4, 10 and 30 keV  are constructed  (see  Fig.~\ref{fig3}).  The experimental measurements were performed  at the Justus-Liebig  University of Giessen.   In Fig.~\ref{fig4}, the X-ray spectra emitted by highly charged ions of  the  cathode material  are presented for different electron beam energies. The radiation of iridium ions in charge states up to $q=+63$ was identified. Analysis of the high-energy part of  spectra due to radiative recombination with  incident electrons allows one to estimate the current density $j_e$ at the level of about 10-20~kA/cm$^2$ \cite{14}.  A comparison of  X-ray spectra from highly charged ions of  cathode material in the classical Livermore EBIT (see Fig.~5 of  work~\cite{5})  and those obtained with the use of the MaMFIS-10 shows that these devices have comparable ionization efficiency.  The first measurements of   X-ray emission spectra from the  MaMFIS-30 were performed for two discrete values of the magnetic field $B_c$ at  cathode and different geometry of  electrodes of the electron gun. Obviously, it is possible to implement a smooth movement of  cathode in the focusing magnetic field and the fine tuning for the distance between cathode and anode in order to provide with sufficiently wide ranges for  energy and current density of the electron beam.

\section{Classification of ion sources with respect to electron beam energy}

The field of applications of EBITs is rather extensive. However, one can distinguish two areas of  research, which require complementary  energy ranges of the electron beam and, accordingly, different designs. The first area of research covers the low-charge ions produced in  low-energy electron beams. In particular, one can mention here the spectroscopic data relevant to the sources of the extreme ultraviolet radiation (ions of Sn and Xe with the charge $q \sim +10$) and diagnostics of  high-temperature plasmas related to the solar corona (ions of  Fe with $q\sim +10$) and to fusion plasmas in  the ITER reactor  (ions of  tungsten with $q \sim +20$).  In 2012,  a compact EBIT  with the energy of electron beam from 100 eV up to  2.5 keV was developed for these purposes at the University of Electro-Communications in Tokyo \cite{8}.  Similar devices were also built at the  Fudan University and  the National Institute  of  Standards and Technology \cite{15,16}.  

From the other side, there is still a long-standing problem of complete ionization of  heavy elements.  In literature, it is known  only one work, where about ten fully stripped U$^{92+}$ ions were produced and trapped  \cite{6}.  Despite long-term research on the SuperEBITs in Tokyo, Heidelberg and Shanghai, no one could  reproduce the results obtained at LLNL.  Since the binding energy of the $K$-shell electrons in uranium is about 130 keV,  the  electron beam for such experiments should have very  high energy  not less than  200--250 keV.  For ionization of  all electron shells except for the $K$ shell  of any heavy element up to uranium, the  energy of  electron beam should be at least 60 keV.

It is natural to separate  a problem of ionization of the $K$-shell electrons in heavy elements into a  particular challenge.  The energy regimes of  electron beam within the range  from 100~eV up to 60~keV can be realized in a single  universal device,  which combines technologies of the EBIS/T  and  MaMFIS. Therefore, it is enough to design just two ion sources for complete ionization of  any element of the periodic table.

\begin{figure}[b]
\begin{minipage}{13.7pc}
\includegraphics[width=13.7pc]{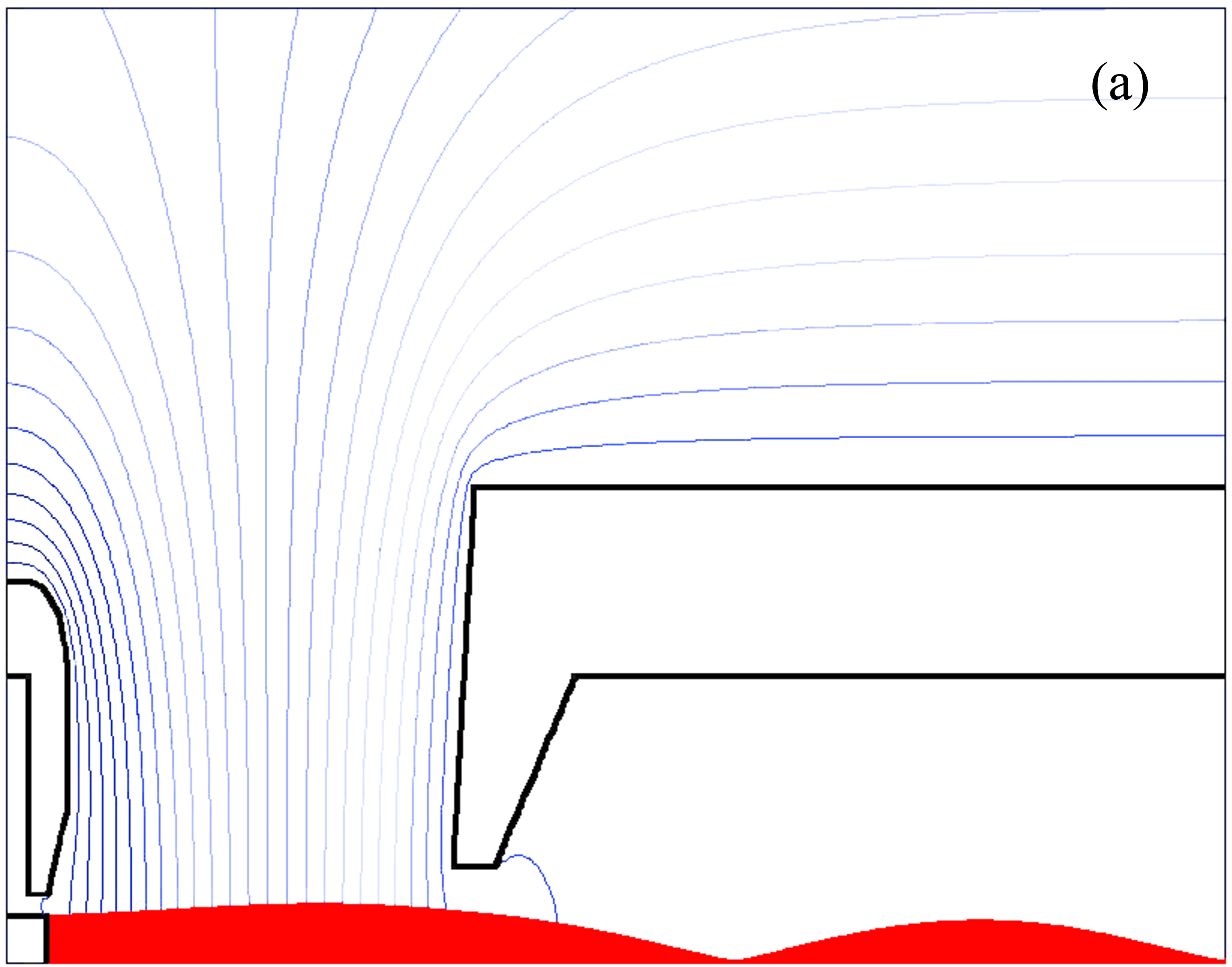}
\end{minipage}\hspace{0.4pc}%
\begin{minipage}{13.7pc}
\includegraphics[width=13.7pc]{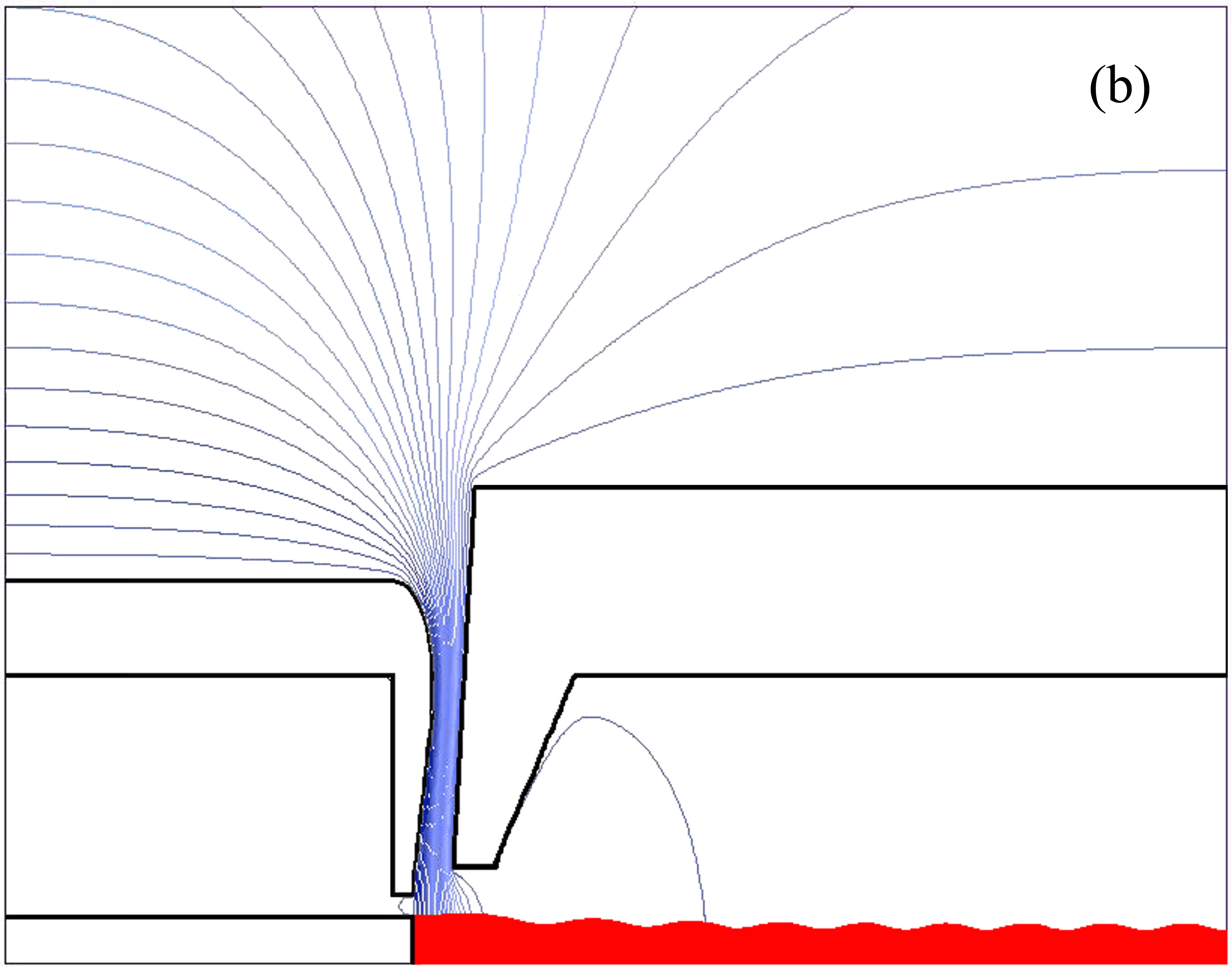}
\end{minipage} \hspace{0.2pc}%
\begin{minipage}{8.7pc}
\includegraphics[width=8.7pc]{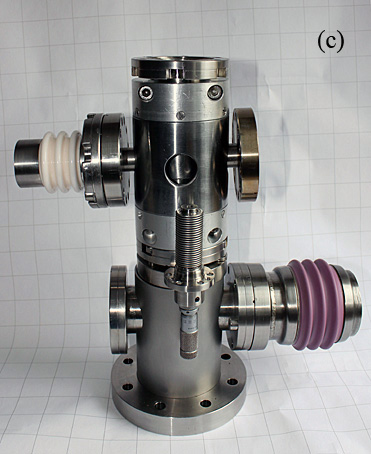}
\end{minipage} 
\caption{\label{fig5}  Electron trajectories for the electron gun with variable cathode-anode distance. (a)~Rippled electron beam with $j_e = 18$~kA/cm$^2$ at the second focus ($I_e = 200$~mA, $E_e = 30$~keV, $B_c = 100$~G), 
(b)~Low ripple electron beam  ($I_e = 20$~mA, $E_e = 1$~keV, $B_c = 2$~kG) and  (c)~General view of  the UniMaMFIS-60.}
\end{figure}

 \section{Universal main magnetic focus  ion source}
 
The ionization of  both light and heavy elements  in a single device is achieved by  combination of two methods of focusing the electron beam, namely, by mix of the immersed  ($B_c \sim 0$)   and  fully immersed gun  ($B_c  \gg 0$) techniques. In the first case, a rippled electron beam with high energy is formed for operation in the MaMFIS regime. In the second case, a low ripple electron beam  with low energy is formed for operation in the EBIS/T regime.  The movement of  cathode in  the electron gun is implemented by using the $z$-manipulator \cite{17}.  The ion source becomes hybrid and is named the universal MaMFIS (UniMaMFIS). In Fig.~\ref{fig5}, the electron trajectories are simulated  for the MaMFIS-30. The  UniMaMFIS upgraded up to $E_e=60$~keV is also presented.  In Table~\ref{tab1}, the project parameters of the device are given. The expected yield of Xe$^{52+}$ is estimated at the level of about 300 ions per second.

\begin{table}[tb]
\caption{\label{tab1} Project parameters of  the UniMaMFIS-60.}
\begin{center}
\begin{tabular}{lllll}
\hline
\multicolumn{1}{c}{Operational regime} &
\multicolumn{1}{c}{$E_e$~(keV)} &
\multicolumn{1}{c}{$I_e$~(mA) } &
\multicolumn{1}{c}{$j_e$~(A/cm$^2$)} &
\multicolumn{1}{c}{$L_\mathrm{trap}$~(cm)} \\   \hline
\multicolumn{1}{c}{EBIS/T} &  0.1-20 & 10-200   &  10-500 & 2  \\
\multicolumn{1}{c}{MaMFIS} &   20-60 &  200 &20000 &  0.1 \\
\hline
\end{tabular} \end{center}
\end{table}

\begin{center}
\begin{table}[h]
\caption{\label{tab2} Cross section $\sigma^+$ for the $1s$ ionization of  U$^{91+}$ and cross section $\sigma_{R}$  for the radiative recombination with  U$^{92+}$  as functions of impact electron energy $E_e$. }
\centering
\begin{tabular}{clllllllll}
\hline
$E_e$~(keV) &  \multicolumn{1}{c}{150}  &  \multicolumn{1}{c}{200} &   \multicolumn{1}{c}{250} &  \multicolumn{1}{c}{300} &  \multicolumn{1}{c}{350} &  \multicolumn{1}{c}{400} &  \multicolumn{1}{c}{450} &  \multicolumn{1}{c}{500} &   \multicolumn{1}{c}{900} \\
\hline
$\sigma^+$~(b) &\hphantom{0}0.64 &\hphantom{0}1.2 &  \hphantom{0}1.6 & \hphantom{0}1.9 &  \hphantom{0}2.1 &  \hphantom{0}2.3 & \hphantom{0}2.4 & 2.5  & 3.1\\
$\sigma_{R}$~(b) &67   & 43   & 30  &  22  & 17  & 13   & 11  & 9.0  & 2.9 \\
\hline
\end{tabular}
\end{table}
\end{center}

\begin{figure}[t]
\begin{minipage}{18pc}
\includegraphics[width=18pc]{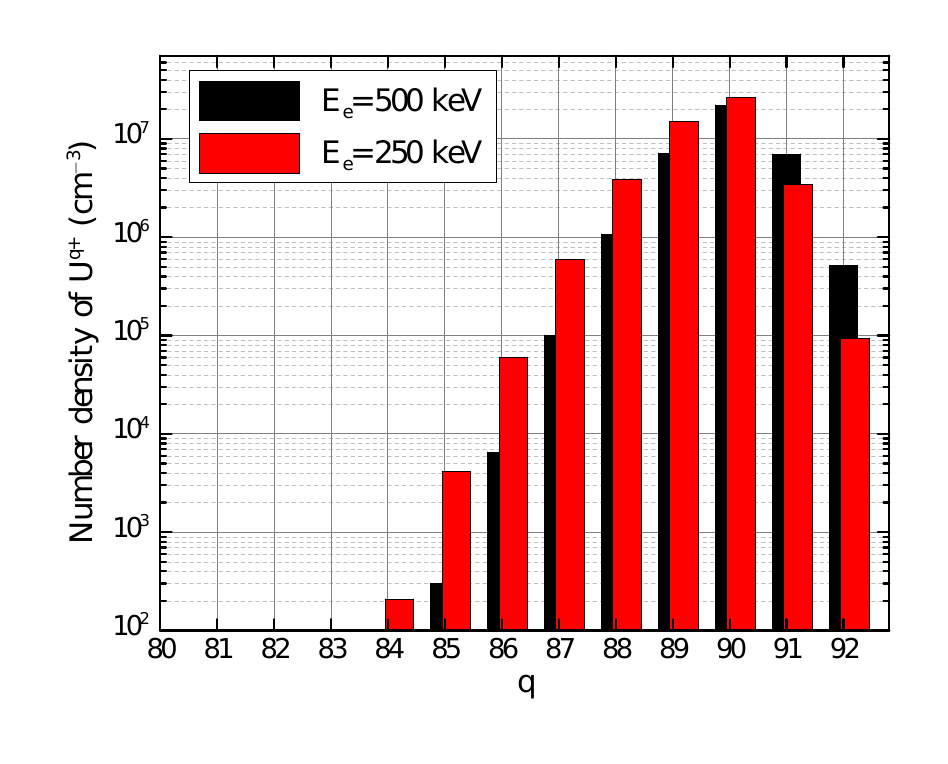}
\caption{\label{fig6} Charge-state distributions of U$^{q+}$ ions for two energies of the electron beam ($I_e= 2.5$~A, $j_e=10$~kA/cm$^2$, cooling by neutral helium at pressure of about $10^{-10}$ mbar).}
\end{minipage}\hspace{2pc}%
\begin{minipage}{18pc}
\includegraphics[width=18pc]{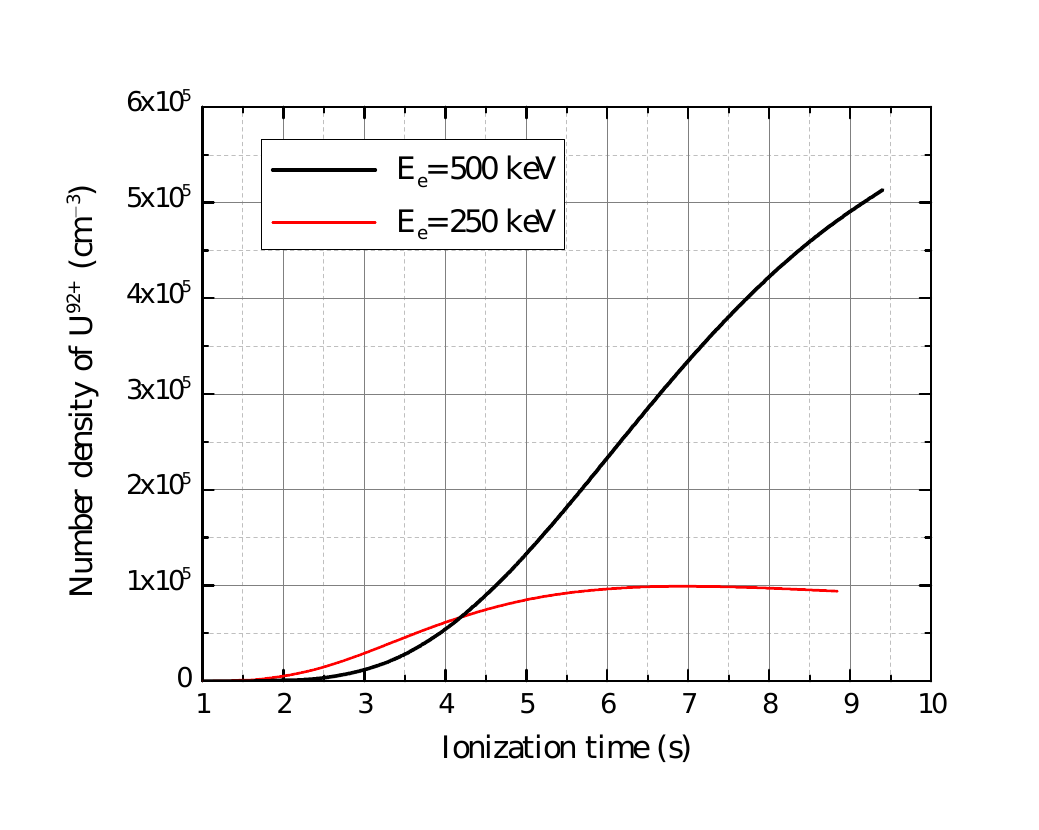}
\caption{\label{fig7} Number density of U$^{92+}$ ions vs ionization time $\tau$ for two energies of the electron beam ($I_e= 2.5$~A, $j_e=10$~kA/cm$^2$,  cooling by neutral helium at pressure of about $10^{-10}$ mbar).}
\end{minipage} 
\end{figure}

\section{Complete ionization of  heavy and superheavy elements}

The extremely high electron current density achieved in the ion trap together with relatively small size of the device and possibility to use the permanent magnets for focusing the electron beam allow one to apply the MaMFIS technology for efficient ionization of uranium and transuranium elements up to bare nuclei. The modern vacuum systems can provide ultra-low pressure (lower than $10^{-9}$~mbar) of  working gas in the MaMFIS chamber at a room temperature. 

A  steady charge-state distribution of  ions  is characterized by equality of the rates of ionization and competing processes. The latter involves mainly charge exchange and radiative recombination.  Since the rate of  charge exchange depends on  number density of the working gas, the corresponding contribution can be suppressed  at the expense of  ultra-high vacuum in the ion trap. The radiative recombination occurs predominantly into the  innermost electron shells, but its cross section $\sigma_R$ decreases with increasing the electron energy  (see Table \ref{tab2}). The K-shell ionization cross section $\sigma^+$ is calculated  for  U$^{91+}$ ions, taking into account the relativistic corrections \cite{18}.  As can be seen, the cross sections  $\sigma_R$ and $\sigma^+$ become comparable at $E_e \sim 900$~keV, that is, at about 7  threshold units. The production of bare uranium at $E_e \sim 200$~keV is not very efficient, since $\sigma_R$ is about 40 times larger than $\sigma^+$.  In Figs.~\ref{fig6} and \ref{fig7}, the number densities of   U$^{q+}$ ions  are given as functions of charge states $q$ and ionization time $\tau$, respectively. The computer simulations were performed according to work \cite{19} for two energies of the electron beam. The corresponding estimates for yield of  bare uranium  give 1300 and 5000 nuclei  per hour at $E_e=250$~keV and 500~keV, respectively.

\section*{Acknowledgements}

The authors are indebted to A.~M\"{u}ller for giving opportunity to test  MaMFIS at the Justus-Liebig University of Giessen, to A.~Borovik~Jr.  for his support in X-ray measurements and to O.K.~Kultashev and  V.~Kogan for their contribution to  simulations of the electronic optics.

%\section*{References}

\end{document}